\documentclass[prd,superscriptaddress,amsfonts,amssymb,amsmath,showpacs,onecolumn]{revtex4-2}

\usepackage{amsmath, amssymb, amsfonts} 
\usepackage{bm} 
\usepackage{graphicx} 
\usepackage{hyperref} 
\usepackage[dvipsnames]{xcolor} 
\usepackage{booktabs} 
\usepackage{dcolumn} 
\usepackage{enumitem} 
\usepackage{mathtools} 
\usepackage{mathrsfs} 
\usepackage{setspace} 
\usepackage{float} 

\usepackage{makecell} 
\usepackage{multirow} 
\usepackage{caption, subcaption} 
\usepackage{commath} 
\usepackage{ragged2e} 
\usepackage{orcidlink} 
\usepackage{mathalfa} 
\usepackage{calligra} 
\usepackage[misc]{ifsym} 

\hypersetup{
    colorlinks=true,
    citecolor=blue,
    linkcolor=red,
    filecolor=magenta,
    urlcolor=blue
}

\allowdisplaybreaks[1] 

\def\jnl@style{}
\def\aaref@jnl#1{{\jnl@style#1}}

\def\aj{\aaref@jnl{AJ}}                   
\def\apj{\aaref@jnl{ApJ}}                 
\def\apjl{\aaref@jnl{ApJ}}                
\def\apjs{\aaref@jnl{ApJS}}               
\def\apss{\aaref@jnl{Ap\&SS}}             
\def\aap{\aaref@jnl{A\&A}}                
\def\aapr{\aaref@jnl{A\&A~Rev.}}          
\def\aaps{\aaref@jnl{A\&AS}}              
\def\mnras{\aaref@jnl{Mon.~Not.~Roy.~Astron.~Soc.}} 
\def\prd{\aaref@jnl{Phys.~Rev.~D}}        
\def\plb{\aaref@jnl{Phys.~Lett.~B}}       
\def\prc{\aaref@jnl{Phys.~Rev.~C}}        
\def\prl{\aaref@jnl{Phys.~Rev.~Lett.}}    
\def\qjras{\aaref@jnl{QJRAS}}             
\def\skytel{\aaref@jnl{S\&T}}             
\def\ssr{\aaref@jnl{Space~Sci.~Rev.}}     
\def\zap{\aaref@jnl{ZAp}}                 
\def\nat{\aaref@jnl{Nature}}              
\def\aplett{\aaref@jnl{Astrophys.~Lett.}} 
\def\apspr{\aaref@jnl{Astrophys.~Space~Phys.~Res.}} 
\def\physrep{\aaref@jnl{Phys.~Rep.}}      
\def\physscr{\aaref@jnl{Phys.~Scr}}       
\def\commat{\aaref@jnl{Comm.~Math.~Phys.}} 
\def\science{\aaref@jnl{Science}}         
\def\cqg{\aaref@jnl{Classical Quant.~Grav.}} 
\def\jpcs{\aaref@jnl{JPCS}}               
\def\ijmpd{\aaref@jnl{Int.~J.~Mod.~Phys.~D}} 
\def\grg{\aaref@jnl{Gen.~Relat.~Gravit.}} 
\def\rpp{\aaref@jnl{Rep.~Prog.~Phys.}}    
\def\npa{\aaref@jnl{Nucl.~Phys.~A}}       
\def\lrr{\aaref@jnl{Living Rev.~Rel.}}    
\def\jcap{\aaref@jnl{J.~Cosmology Astropart.~Phys.}} 
\def\rmp{\aaref@jnl{Rev.~Mod.~Phys.}}     
\def\epjc{\aaref@jnl{Eur.~Phys.~J.~C}}    

\begin{document}

\preprint{APS/123-QED}

\title{Traversable Wormhole Solutions in $f(\mathscr{Q},\mathscr{L}_m)$ Gravity } %

\author{K. Suhasini\orcidlink{0009-0000-0591-5590}}%
\email{suhasini.kandukoori25@gmail.com}%
\affiliation{Department of Mathematics, 
 SR University, Ananthasagar, Hasanparthy, Warangal - 506371, Telangana, India}

\author{G. Ravi Kiran\orcidlink{0000-0002-5496-3879}}
\email{ravikiran.wgl@gmail.com}
\affiliation{Department of Mathematics, 
 SR University, Ananthasagar, Hasanparthy, Warangal - 506371, Telangana, India}%
 
\author{N. S. Kavya\orcidlink{0000-0001-8561-130X}}
\email{kavya.samak.10@gmail.com, kavya.ns@christuniversity.in}
\affiliation{Centre for Mathematical Needs, Department of Mathematics, CHRIST (Deemed to be University), Bengaluru 560029, India.}%

\author{C. S. Varsha\orcidlink{0009-0008-3529-0390}}
\email{varshaagowda0902@gmail.com}
\affiliation{Department of P.G. Studies and Research in Mathematics,
 Kuvempu University, Shankaraghatta, Shivamogga 577451, Karnataka, INDIA}%

\author{V. Venkatesha\orcidlink{0000-0002-2799-2535}}%
\email{vensmath@gmail.com}
\affiliation{Department of P.G. Studies and Research in Mathematics,
 Kuvempu University, Shankaraghatta, Shivamogga 577451, Karnataka, INDIA}%

\date{\today}

\begin{abstract}
We investigate traversable wormhole solutions within the framework of $f(\mathscr{Q},\mathscr{L}_m)$ gravity, a symmetric teleparallel theory featuring non-minimal coupling between geometry and matter. Adopting a linear functional form $f(\mathscr{Q},\mathscr{L}_m) = -\alpha \mathscr{Q} + 2\mathscr{L}_m + \beta$, we derive the field equations for a static, spherically symmetric Morris-Thorne wormhole geometry with vanishing redshift function. Four distinct shape functions are considered: $b(r)=\sqrt{r_0 r}$, $b(r)=r_0\left(\dfrac{r}{r_0}\right)^{\gamma}$ (with $0<\gamma<1$), and $b(r)=\dfrac{r_0 \ln (r+1)}{\ln (r_0+1)}$. The geometric viability of each configuration is verified through standard traversability conditions, including the flaring-out requirement and asymptotic flatness. We analyze the energy conditions and demonstrate that, consistent with known results in wormhole physics, the null energy condition is violated in the vicinity of the throat, indicating the presence of exotic matter. In addition, we employ embedding diagrams to visualize the spatial geometry of the wormhole solutions, providing a clear geometric interpretation of the flaring-out condition at the throat.
Our results suggest that $f(\mathscr{Q},\mathscr{L}_m)$ gravity provides a viable framework for constructing traversable wormholes, with the non-minimal matter-geometry coupling influencing both the geometry and the matter sector.
\end{abstract}

\keywords{wormholes, modified gravity, $f(\mathscr{Q},\mathscr{L}_m)$ gravity, symmetric teleparallel gravity, energy conditions}%

\maketitle

\tableofcontents

\section{Introduction}\label{section1}

A wormhole is a hypothetical tunnel-like structure connecting two distant regions of spacetime, offering a potential shortcut for intergalactic travel. These geometries arise as solutions in Einstein's general relativity and its extensions, representing one of the theory's most intriguing predictions. The concept of traversable wormholes was rigorously formulated by Morris and Thorne \cite{morris1988wormholes}, who demonstrated that such structures could, in principle, allow safe passage between distant cosmic regions. Visser later provided a comprehensive analysis of Lorentzian wormholes and their physical properties \cite{Visser:1995cc}. These foundational studies revealed that classical wormhole solutions in general relativity inevitably require exotic matter that violates the null energy condition, presenting a significant theoretical challenge.

Various approaches have been proposed to mitigate or physically justify this exotic matter content, including specific choices of shape functions \cite{Kuhfittig:1999nd,Kuhfittig:2005gy}, phantom energy sources \cite{Armendariz-Picon:2002gjc,Sushkov:2005kj, Rahaman:2013xoa}, and configurations admitting conformal symmetries \cite{Boehmer:2007md,Boehmer:2007rm}. Beyond purely theoretical interest, wormholes have also been investigated from phenomenological and observational perspectives. Traversability conditions and quantum effects have been explored in different contexts \cite{Gao:2016bin,Nicolis:2009qm}, while gravitational lensing signatures and optical properties have been proposed as potential observational probes \cite{Ovgun:2018fnk,Kuhfittig:2013hva, Varsha:2024iwx}. The construction of dynamical and thin-shell wormholes further enriches the landscape of viable geometries \cite{Lobo:2020kxn,Rahaman:2010bt,Rahaman:2006vg}.

In recent decades, modified theories of gravity have emerged as promising frameworks for constructing wormhole solutions with relaxed or reconceptualized energy condition requirements. Early works in \( f(R) \) gravity demonstrated that higher-order curvature terms could effectively support wormhole geometries \cite{Lobo:2009ip,Harko:2013yb,Hassan:2021egb, Kavya:2025ygq, Rahaman:2013qza, Jamil:2013tva, Venkatesha:2023zkm, Kavya:2023tjf, Kavya:2023mwv, Varsha:2025qgv}. This theme was subsequently extended to curvature–matter coupling models such as \( f(R,T) \) gravity \cite{Zubair:2016cde,Banerjee:2019wjj,Bhar:2023ihi, Kavya:2022dam, Kavya:2025glb}. Wormhole solutions have also been studied in other gravitational frameworks, including Einstein–Gauss–Bonnet theory \cite{Kanti:2011jz}, dilaton–axion gravity \cite{Usmani:2010cd}, Rastall gravity \cite{Mustafa:2019oiy}, and Finslerian geometry \cite{SinghKsh:2023fphy}.

A compelling alternative to curvature-based descriptions arises within the teleparallel framework, where gravity is attributed to torsion or non-metricity rather than curvature. The symmetric teleparallel equivalent of general relativity (STEGR) provides a geometrically distinct but dynamically equivalent description of Einstein's theory \cite{BeltranJimenez:2017tkd}. In this formulation, the gravitational interaction is encoded in the non-metricity scalar \( Q \), leading naturally to \( f(Q) \) theories as extensions. Theoretical reconstruction schemes and observational constraints have shown that \( f(Q) \) gravity can successfully account for late-time cosmic acceleration \cite{Capozziello:2022wgl,Lazkoz:2019sjl,Anagnostopoulos:2021ydo, Mishra:2024shg,Mishra:2024oln,Kavya:2024ssu,Naik:2023ykt}.

Motivated by these successes, wormhole geometries in symmetric teleparallel gravity have begun to attract attention. Static and spherically symmetric solutions, together with their energy conditions, have been explored in \( f(Q) \) gravity \cite{Mustafa:2021ykn,Tayde:2022lxd,Banerjee:2021mqk,Kavya:2024bpj}. The inclusion of quantum effects such as Casimir energy has further enriched this framework, leading to viable Casimir-supported wormholes in \( f(Q) \) gravity \cite{Hassan:2022hcb,Varsha:2025oik}. These developments suggest that non-metricity-based theories offer fertile ground for constructing physically interesting wormhole solutions.

An important generalization within the teleparallel paradigm is the introduction of explicit non-minimal coupling between geometry and matter. This leads to theories such as \( f(Q,T) \) gravity \cite{Xu:2019sbp} and, more fundamentally, \( f(\mathscr{Q},\mathscr{L}_m) \) gravity, where the gravitational Lagrangian depends on both the non-metricity scalar \( \mathscr{Q} \) and the matter Lagrangian density \( \mathscr{L}_m \). Such couplings naturally induce non-conservation of the energy–momentum tensor and modify the effective gravitational dynamics near matter sources. While \( f(Q,T) \) gravity has been studied in various contexts, including cosmology \cite{Najera:2021puo,Shiravand:2022ccb, Chalavadi:2023zcw, Venkatesha:2023tay, Moraes:2019pao, Venkatesha:2025cbl,Kavya:2023lms,venkatesha2025noncommutative} and compact objects, the \( f(\mathscr{Q},\mathscr{L}_m) \) formulation—with its direct coupling to \( \mathscr{L}_m \)—remains relatively unexplored for strong-field geometries. The gravitational framework of \( f(\mathscr{Q},\mathscr{L}_m) \) gravity was first introduced by Hazarika et al. in late 2024 \cite{Hazarika:2024alm}, extending symmetric teleparallel gravity through an explicit matter–geometry coupling; in their foundational work, the authors formulated the theory, derived the corresponding field equations, and explored its cosmological implications. In the years following the initial proposal, the theoretical structure and cosmological consequences of \( f(\mathscr{Q},\mathscr{L}_m) \)gravity have been further investigated in several works. For example, Myrzakulov et al. studied the energy conditions of different \( f(\mathscr{Q},\mathscr{L}_m) \) models within the FLRW framework \cite{Myrzakulov:2025yxy}, and explored observational constraints and bulk-viscous cosmologies in this theory \cite{Myrzakulov:2025ycj,Myrzakulov:2024kuu}. Other recent analyses include examinations of gravitational baryogenesis in specific \( f(\mathscr{Q},\mathscr{L}_m) \) forms \cite{Samaddar:2024qno,Samaddar:2025qxr,Sharif:2025rko}.

This work presents the first investigation of traversable wormhole solutions in \( f(\mathscr{Q},\mathscr{L}_m) \) gravity, filling a notable gap in the literature. We derive the field equations for a static, spherically symmetric Morris–Thorne wormhole and examine several shape functions to assess geometric viability and energy condition profiles within this novel theoretical framework.

This paper is organized as follows. Section~\ref{section2} outlines the theoretical foundations of $f(\mathscr{Q},\mathscr{L}_m)$ gravity, including the field equations and the non-conservation of the energy--momentum tensor. Section~\ref{section3} presents the wormhole metric, derives the corresponding field equations, and introduces the specific shape functions considered. Section~\ref{section4} analyzes the energy conditions for each solution. Section~\ref{sectionemb} is devoted to an explicit embedding analysis of the chosen shape functions. Finally, Section~\ref{section5} provides a discussion of the results, and Section~\ref{section6} presents our conclusions and outlines possible directions for future research.

\section{Theoretical Framework of $f(\mathscr{Q},\mathscr{L}_m)$ Gravity}\label{section2}

General relativity traditionally attributes gravitational interaction to spacetime curvature, described by the Riemann tensor. However, equivalent geometric formulations exist, including the Teleparallel Equivalent of General Relativity (where gravity arises from torsion) and the Symmetric Teleparallel Equivalent of General Relativity (where gravity is governed by non-metricity). In symmetric teleparallel gravity, the gravitational interaction is entirely encoded in the non-metricity scalar $\mathscr{Q}$, which quantifies the deviation of the metric tensor $g_{\mu\nu}$ from being covariantly constant ($\nabla_{\lambda} g_{\mu\nu} \neq 0$). This formulation employs a torsion-free, non-metric connection $\Gamma^\lambda_{\mu\nu}$.

We investigate $f(\mathscr{Q},\mathscr{L}_m)$ gravity, which generalizes the symmetric teleparallel equivalent of general relativity by introducing a gravitational Lagrangian that depends arbitrarily on both the non-metricity scalar $\mathscr{Q}$ and the matter Lagrangian $\mathscr{L}_m$. The action for this theory is defined as:
\begin{equation}
S = \int f(\mathscr{Q},\mathscr{L}_m)\sqrt{-g}\, d^4x,
\label{action}
\end{equation}
where $g$ denotes the determinant of the metric tensor $g_{\mu\nu}$. This non-minimal coupling between geometry and matter provides a rich framework for addressing cosmological puzzles and may yield observable signatures distinct from the $\Lambda$CDM model.

\subsection{Non-metricity and Disformation}

The non-metricity tensor measures the failure of the metric to be covariantly conserved and is defined as:
\begin{equation}
Q_{\gamma\mu\nu} = -\nabla_{\gamma} g_{\mu\nu}
= -\partial_{\gamma} g_{\mu\nu} + g_{\nu\sigma}\, \tilde{\Gamma}^\sigma_{\mu\gamma} + g_{\sigma\mu}\, \tilde{\Gamma}^\sigma_{\nu\gamma},
\label{Qtensor}
\end{equation}
where $\tilde{\Gamma}^\sigma_{\mu\nu}$ denotes the symmetric teleparallel (Weyl) connection. The associated disformation tensor $L^\beta_{\ \alpha\gamma}$ is given by:
\begin{equation}
L^\beta_{\ \alpha\gamma} = \frac{1}{2} g^{\beta\eta}
\left(Q_{\gamma\alpha\eta} + Q_{\alpha\eta\gamma} - Q_{\eta\alpha\gamma}\right).
\label{disformation}
\end{equation}

The non-metricity scalar is constructed as \cite{BeltranJimenez:2017tkd}:
\begin{equation}
\mathscr{Q} = -g^{\mu\nu}\left(L^\beta_{\ \alpha\mu} L^\alpha_{\ \nu\beta} - L^\beta_{\ \alpha\beta} L^\alpha_{\ \mu\nu}\right).
\label{Qscalar}
\end{equation}

\subsection{Superpotential (Non-metricity Conjugate)}

The superpotential tensor (or non-metricity conjugate) is defined as:
\begin{equation}
P^{\beta}{}_{\mu\nu} \equiv \frac{1}{4}
\left[-Q^{\beta}{}_{\mu\nu} + 2 Q_{(\mu\nu)}^{\ \ \ \beta}
+ Q^{\beta} g_{\mu\nu} - \tilde{Q}^{\beta} g_{\mu\nu}
- \delta^{\beta}_{(\mu} Q_{\nu)}\right],
\label{Pdef}
\end{equation}
where the traces of non-metricity are:
\begin{equation}
Q_{\beta} = Q_{\beta\mu}{}^{\mu},
\qquad
\tilde{Q}_{\beta} = Q_{\mu\beta}{}^{\mu}.
\label{Qtraces}
\end{equation}
The scalar $\mathscr{Q}$ can equivalently be expressed as:
\begin{equation}
\mathscr{Q} = - Q_{\beta\mu\nu} P^{\beta\mu\nu}
= -\frac{1}{4}\left(-Q_{\beta\nu\rho} Q^{\beta\rho\nu}
+ 2 Q_{\beta\rho\nu} Q^{\rho\beta\nu} - 2 Q_{\rho} \tilde{Q}^{\rho}
+ Q_{\rho} Q^{\rho}\right).
\label{Qalt}
\end{equation}

\subsection{Field Equations}

Varying the action~\eqref{action} with respect to the metric tensor $g_{\mu\nu}$ yields the gravitational field equations:
\begin{equation}
\frac{2}{\sqrt{-g}} \nabla_{\alpha}\!\left(f_\mathscr{Q}\,\sqrt{-g} P^{\alpha}{}_{\mu\nu}\right)
+ f_\mathscr{Q}\left(P_{\mu}{}^{\alpha\beta} Q_{\nu\alpha\beta}
- 2 Q_{\alpha\beta\mu} P^{\alpha\beta}{}_{\nu}\right)
+ \frac{1}{2} f\, g_{\mu\nu}
= \frac{1}{2} f_{\mathscr{L}_m} \left(g_{\mu\nu} \mathscr{L}_m - T_{\mu\nu}\right),
\label{Fieldeq}
\end{equation}
where:
\begin{equation}
f_\mathscr{Q}\equiv \frac{\partial f(\mathscr{Q},\mathscr{L}_m)}{\partial \mathscr{Q}},
\qquad
f_{\mathscr{L}_m} \equiv \frac{\partial f(\mathscr{Q},\mathscr{L}_m)}{\partial \mathscr{L}_m},
\label{fdefs}
\end{equation}
and the energy-momentum tensor of matter is:
\begin{equation}
T_{\mu\nu} = g_{\mu\nu} \mathscr{L}_m - 2 \frac{\partial \mathscr{L}_m}{\partial g^{\mu\nu}}.
\label{Tmunu}
\end{equation}

Variation with respect to the affine connection gives an additional relation:
\begin{equation}
\nabla_\mu \nabla_\nu \left(4 \sqrt{-g}f_\mathscr{Q}P^{\mu\nu}{}_{\alpha} + H^{\mu\nu}{}_{\alpha}\right) = 0,
\label{ConnVar}
\end{equation}
where $H^{\mu\nu}{}_{\alpha}$ denotes the hypermomentum density:
\begin{equation}
H^{\mu\nu}{}_{\alpha} = \sqrt{-g}f_{\mathscr{L}_m} \frac{\delta \mathscr{L}_m}{\delta Y^{\alpha}{}_{\mu\nu}}.
\label{Hdef}
\end{equation}

\subsection{Energy-Momentum Non-Conservation}

Taking the covariant divergence of Equation~\eqref{Fieldeq} leads to:
\begin{equation}
\nabla_\mu T^{\mu}{}_{\nu} = B_{\nu} \neq 0,
\label{NonCons}
\end{equation}
where the term $B_{\nu}$ represents energy-momentum exchange between geometry and matter sectors. This non-conservation arises explicitly from the coupling between $\mathscr{Q}$ and $\mathscr{L}_m$ in the function $f(\mathscr{Q},\mathscr{L}_m)$.

For the special case $f(\mathscr{Q},\mathscr{L}_m)= f(\mathscr{Q}) + 2\mathscr{L}_m$, the field equations reduce to those of standard $f(\mathscr{Q})$ gravity, and the energy-momentum tensor is conserved:
\begin{equation}
\nabla_\mu T^{\mu}{}_{\nu} = 0.
\label{Conservation}
\end{equation}

\section{Traversable Wormhole Configurations in $f(\mathscr{Q},\mathscr{L}_m)$ Gravity}\label{section3}

We now investigate traversable wormhole geometries within $f(\mathscr{Q},\mathscr{L}_m)$ gravity. The spacetime is described by the static, spherically symmetric Morris-Thorne wormhole metric:
\begin{equation} 
ds^{2} = e^{2\Phi(r)} dt^{2} - \left(1 - \frac{b(r)}{r}\right)^{-1} dr^{2} - r^{2} (d\theta^{2} + \sin^{2}\theta\, d\phi^{2}),
\label{wormhole_metric}
\end{equation}
where $\Phi(r)$ is the redshift function and $b(r)$ is the shape function. The throat is located at $r = r_0$, where $b(r_0) = r_0$. Traversability requires the flaring-out condition $b'(r_0) < 1$ and the absence of event horizons, which is ensured by $1 - b(r)/r > 0$.

The matter content threading the wormhole is modelled as an effective isotropic fluid with an energy–momentum tensor
\begin{equation}
T^{\mu}{}_{\nu} = (\rho + p) u^{\mu} u_{\nu} - p \delta^{\mu}_{\nu},
\label{perfectfluid_tensor}
\end{equation}
where $\rho$ and $p$ represent the effective energy density and pressure, respectively. It is important to emphasize that, in $f(\mathscr{Q},\mathscr{L}_m)$ gravity, the non-minimal coupling between geometry and matter modifies the gravitational field equations in such a way that the effective stress–energy contributions governing the spacetime geometry need not correspond to a fundamental anisotropic fluid. Even when the matter sector is taken to be isotropic, the coupling-induced terms introduce effective anisotropic stresses in the geometric sector, which are sufficient to support traversable wormhole configurations..

We adopt a linear model
\begin{equation}
f(\mathscr{Q},\mathscr{L}_m) = -\alpha \mathscr{Q} + 2\mathscr{L}_m + \beta,
\end{equation}
as a minimal extension of symmetric teleparallel gravity, where $\alpha$ and $\beta$ are coupling constants. The choice guarantees recovery of the symmetric teleparallel equivalent of general relativity for $\alpha = 1$ and $\beta = 0$, while the constant $\beta$ effectively acts as a cosmological constant. Nonlinear terms are deferred to future work to clearly isolate the role of matter--geometry coupling in supporting traversable wormhole configurations.

 For the metric~\eqref{wormhole_metric} with $\Phi(r)=0$, the non-metricity scalar becomes:
\begin{equation}
\mathscr{Q} = -\frac{b(r)}{r^3}\left[b'(r) + \frac{b(r)}{r}\right].
\end{equation}

The field equations yield the following expressions for energy density and pressure:
\begin{equation}
\begin{split}
\rho = &-\frac{1}{2 r^3 (r-b(r))^2}\left[r \sqrt{r^3 (b(r)-r) \left(-8 \alpha  r (b(r)+r) b'(r)+b(r) \left(8 \alpha  b(r)+(1-8 \beta ) r^3+8 \alpha  r\right)+(8 \beta -1) r^4\right)}\right.\\
&\left.-b(r) \left(10 \alpha  r^2 b'(r)+\sqrt{-r^3 (r-b(r)) \left(-8 \alpha  r (b(r)+r) b'(r)+b(r) \left(8 \alpha  b(r)-8 \beta  r^3+r^3+8 \alpha  r\right)+(8 \beta -1) r^4\right)}\right.\right.\\
&\left.\left.+r b(r)^2 \left(10 \alpha +6 \alpha  b'(r)+(2 \beta +1) r^2\right)+(4 \beta +2) r^4\right)-6 \alpha  b(r)^3+(2 \beta +1) r^5\right],
\end{split}
\end{equation}
\begin{equation}
p = -\frac{\sqrt{-r^3 (r-b(r)) \left(-8 \alpha  r (b(r)+r) b'(r)+b(r) \left(8 \alpha  b(r)-8 \beta  r^3+r^3+8 \alpha  r\right)+(8 \beta -1) r^4\right)}-r^3 b(r)+r^4}{4 r^3 (r-b(r))}.
\end{equation}

We consider three distinct shape functions:
\begin{itemize}
\item $b_1(r)=\sqrt{r_0\,r}$,
\item $b_2(r)=r_0\left(\dfrac{r}{r_0}\right)^{\gamma}$ with $0<\gamma<1$,
\item $b_3(r)=\dfrac{r_0 \ln (r+1)}{\ln (r_0+1)}$.
\end{itemize}

\begin{figure*}[htbp]
\centering
\subfloat[$b(r)<r$]{\includegraphics[width=0.45\linewidth]{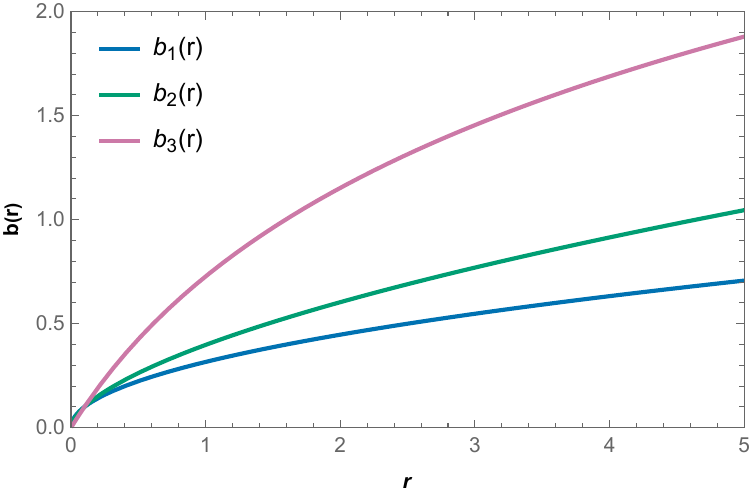}}
\subfloat[$b'(r)<1$]{\includegraphics[width=0.45\linewidth]{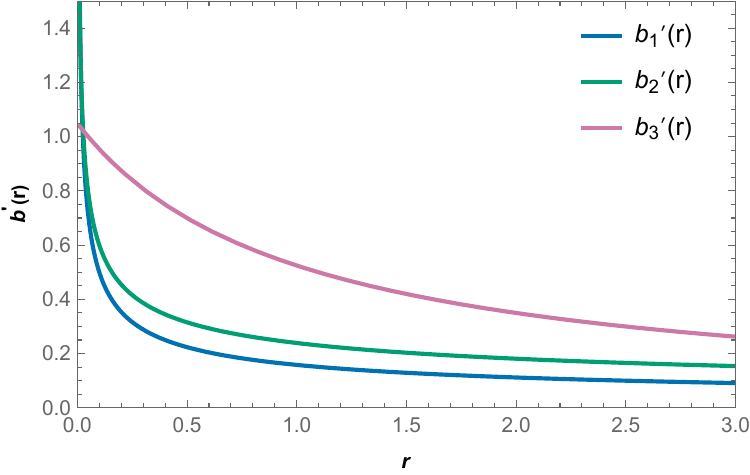}}\\
\subfloat[$\dfrac{b(r)-rb'(r)}{b(r)^2}>0$]{\includegraphics[width=0.45\linewidth]{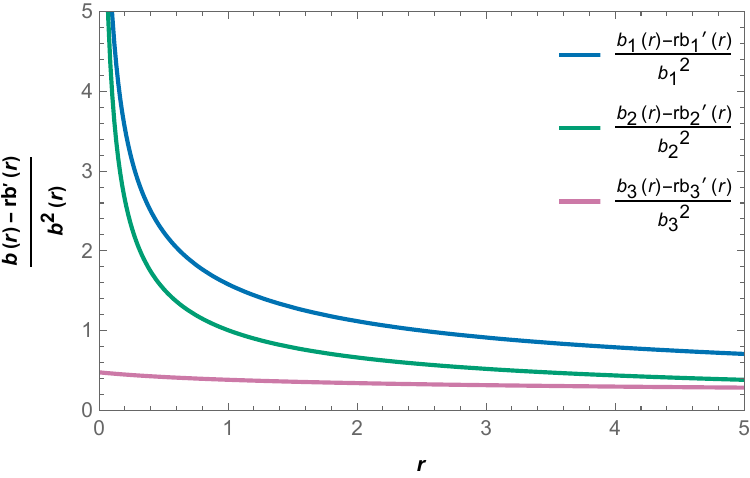}}
\subfloat[$\dfrac{b(r)}{r} \rightarrow 0$ as $r \rightarrow \infty$]{\includegraphics[width=0.45\linewidth]{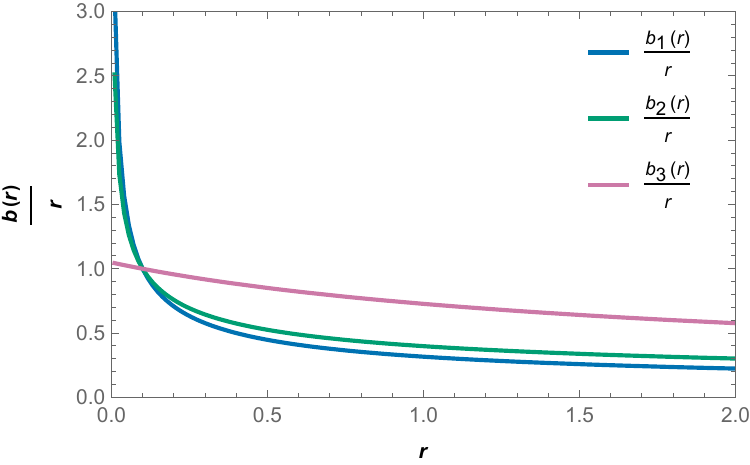}}\\
\subfloat[$b(r)-r<0$]{\includegraphics[width=0.45\linewidth]{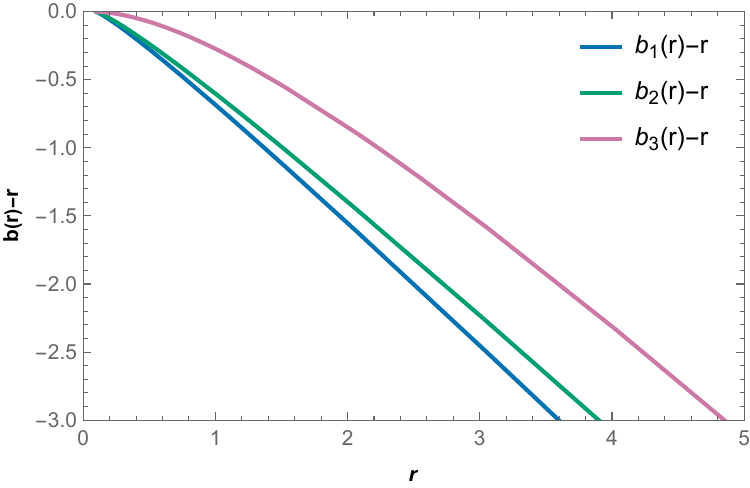}}
\subfloat[$1-\dfrac{b(r)}{r}>0$]{\includegraphics[width=0.45\linewidth]{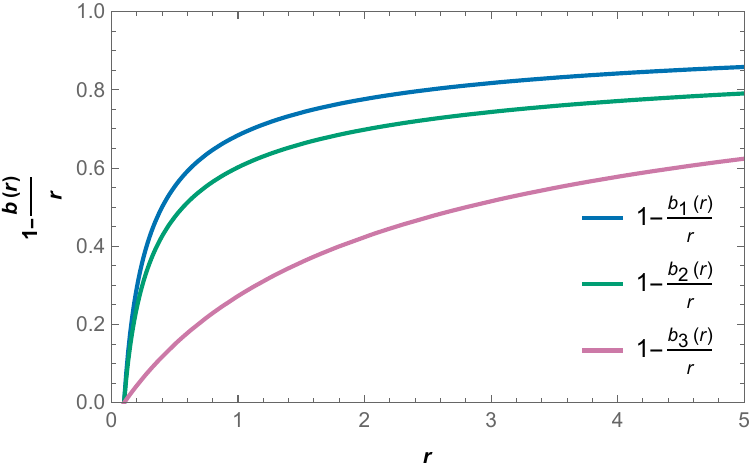}}
\caption{Geometric constraints for wormhole shape functions. (a) $b(r)<r$ ensures a proper throat. (b) $b'(r)<1$ satisfies the flaring-out condition. (c) Positivity of $\frac{b(r)-rb'(r)}{b(r)^2}$ validates traversability. (d) Asymptotic flatness $\frac{b(r)}{r}\to0$. (e) $b(r)-r<0$ reinforces the throat condition. (f) $1-\frac{b(r)}{r}>0$ ensures horizon absence. Different colors correspond to different shape functions.}
\label{fig:shape_conditions}
\end{figure*}

Figure~\ref{fig:shape_conditions} demonstrates that all considered shape functions satisfy the essential geometric requirements for traversable wormholes. The conditions $b(r) < r$ and $b'(r) < 1$ ensure a well-defined throat and flaring-out geometry. The asymptotic behavior $b(r)/r \rightarrow 0$ as $r \rightarrow \infty$ confirms asymptotic flatness, while $1-b(r)/r>0$ guarantees the absence of event horizons.

\section{Energy Conditions}\label{section4}

Energy conditions, derived from the Raychaudhuri equation, impose fundamental constraints on matter and are essential for assessing spacetime viability. For an isotropic fluid with energy density $\rho$ and pressure $p$, the standard energy conditions are:

\begin{itemize}
\item \textbf{Null Energy Condition (NEC):} $\rho + p \geq 0$
\item \textbf{Weak Energy Condition (WEC):} $\rho \geq 0$ and $\rho + p \geq 0$
\item \textbf{Strong Energy Condition (SEC):} $\rho + 3p \geq 0$ and $\rho + p \geq 0$
\item \textbf{Dominant Energy Condition (DEC):} $\rho \geq 0$ and $\rho - |p| \geq 0$
\end{itemize}

A key result in wormhole physics is that traversable wormholes necessarily violate the NEC near the throat, indicating the presence of exotic matter.

\begin{figure*}[htbp]
\centering
\subfloat[$\rho(r)$]{\includegraphics[width=0.45\linewidth]{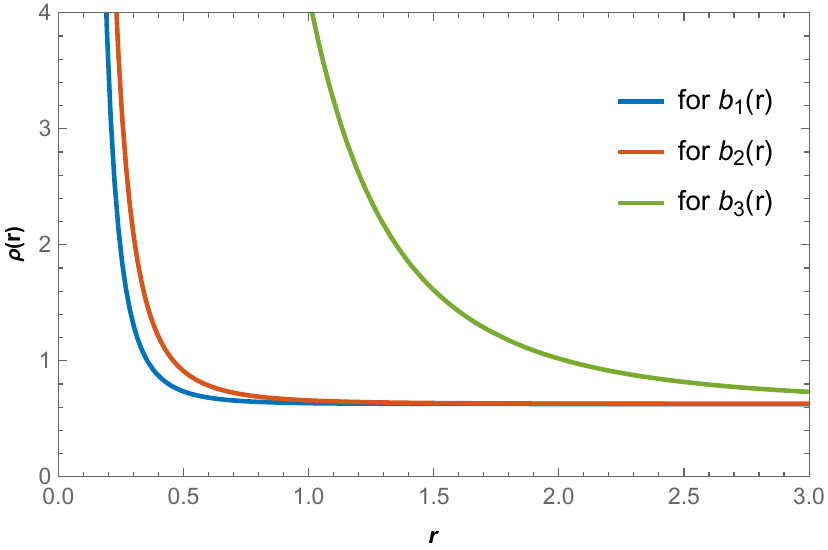}}
\subfloat[$\rho(r) + p(r)$]{\includegraphics[width=0.45\linewidth]{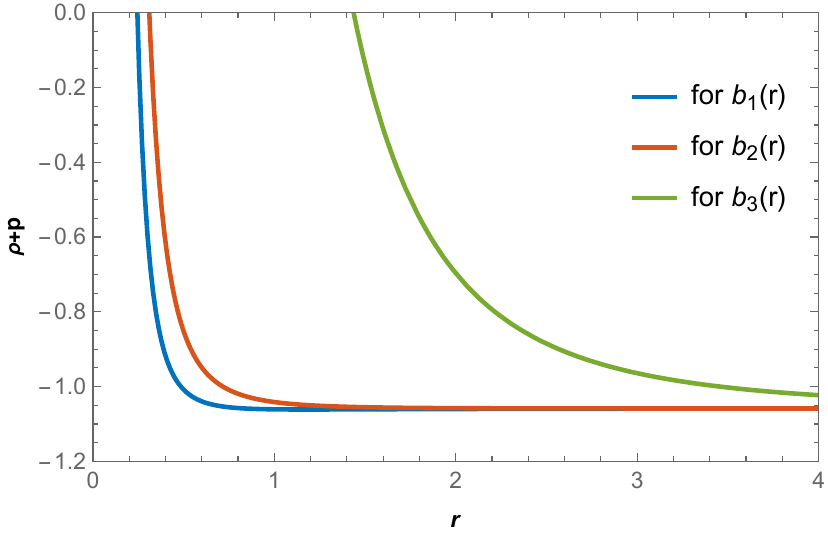}}\\
\subfloat[$\rho(r) - p(r)$]{\includegraphics[width=0.45\linewidth]{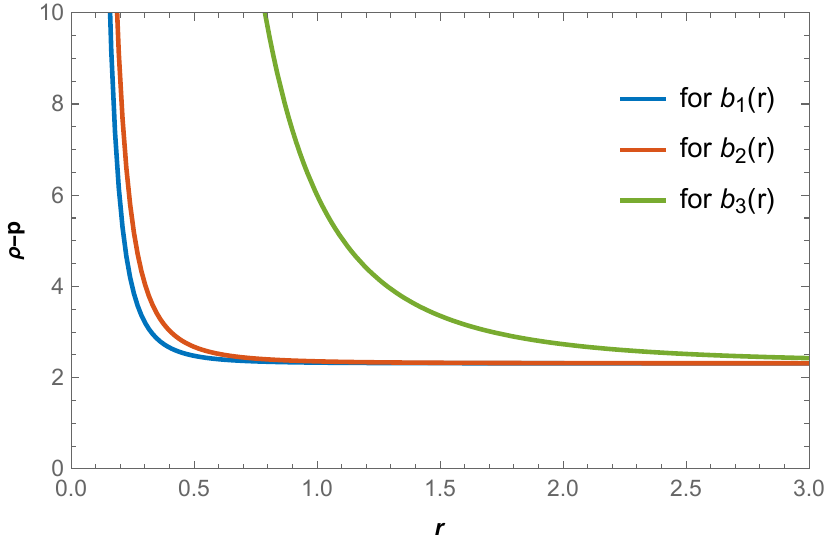}}
\subfloat[$\rho(r) + 3p(r)$]{\includegraphics[width=0.45\linewidth]{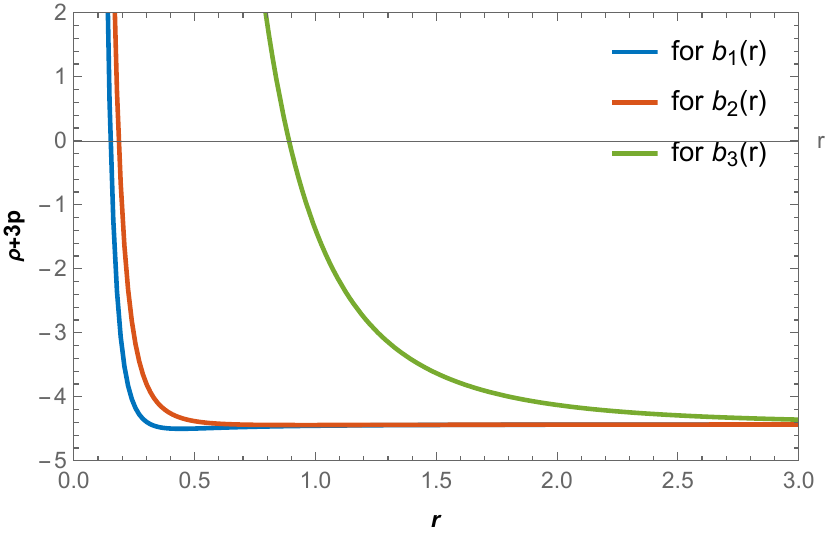}}
\caption{Energy condition profiles for different wormhole shape functions. (a) Energy density $\rho(r)$. (b) NEC term $\rho(r)+p(r)$. (c) DEC term $\rho(r)-p(r)$. (d) SEC term $\rho(r)+3p(r)$. The violation of NEC near the throat is evident, confirming the need for exotic matter.}
\label{fig:energy_conditions}
\end{figure*}

Figure~\ref{fig:energy_conditions} displays the energy condition profiles for our wormhole solutions. As expected for traversable wormholes, the NEC is violated in the vicinity of the throat ($\rho + p < 0$), confirming the presence of exotic matter. The other energy conditions show varied behavior depending on the specific shape function and radial coordinate.

A closer inspection of the null energy condition violation reveals that its magnitude is strongly localized around the wormhole throat and rapidly decreases as the radial coordinate increases. In particular, for all considered shape functions, the quantity $\rho + p$ attains its most negative values in the immediate vicinity of $r = r_0$, while approaching zero at larger distances, indicating that the exotic matter content is confined to a finite region near the throat. Among the examined configurations, the square-root shape function $b(r)=\sqrt{r_0 r}$ exhibits the least severe violation of the NEC, suggesting that it provides the most efficient geometric support for a traversable wormhole within the present framework. Compared to general relativity and standard $f(\mathscr{Q})$ gravity, where stronger or more widespread energy condition violations are typically required, the non-minimal matter--geometry coupling in $f(\mathscr{Q},\mathscr{L}_m)$ gravity effectively redistributes the exoticity into the geometric sector, thereby reducing the dependence on fundamentally exotic matter sources.

\section{Embedding Diagram and Geometric Interpretation of the Shape Function}\label{sectionemb}

To obtain a clearer geometric understanding of the wormhole configurations supported by the chosen shape functions, we analyze their embedding properties. Embedding diagrams provide a powerful visualization tool by representing a two-dimensional spatial slice of the wormhole spacetime in a higher-dimensional Euclidean space. This approach allows one to explicitly verify the flaring-out behavior at the throat, which is a defining requirement for traversable wormholes.

We consider a constant-time $(t=\mathrm{const})$ and equatorial $(\theta=\pi/2)$ slice of the Morris--Thorne wormhole metric given in Equation \eqref{wormhole_metric}, which reduces to
\begin{equation}
ds^2 = \frac{dr^2}{1 - \dfrac{b(r)}{r}} + r^2 d\phi^2 .
\end{equation}
This two-dimensional geometry can be embedded in a three-dimensional Euclidean space described by the line element
\begin{equation}
ds^2 = dz^2 + dr^2 + r^2 d\phi^2 ,
\end{equation}
where $z=z(r)$ characterizes the embedding surface. Equating the above line elements yields the embedding equation
\begin{equation}
\frac{dz}{dr} = \pm \left( \frac{r}{b(r)} - 1 \right)^{-1/2}.
\end{equation}

The behavior of the embedding surface near the throat $r=r_0$, where $b(r_0)=r_0$, determines whether the geometry flares outward. The flaring-out condition is expressed as
\begin{equation}
\left.\frac{d^2 r}{dz^2}\right|_{r=r_0} = \frac{b(r_0)-r_0 b'(r_0)}{2\,b^2(r_0)} > 0 ,
\end{equation}
which requires $b'(r_0) < 1$. This condition ensures that the embedding surface opens outward at the throat, preventing the geometry from pinching off.

\begin{figure*}[htbp]
\centering
\subfloat[]{\includegraphics[width=0.45\linewidth]{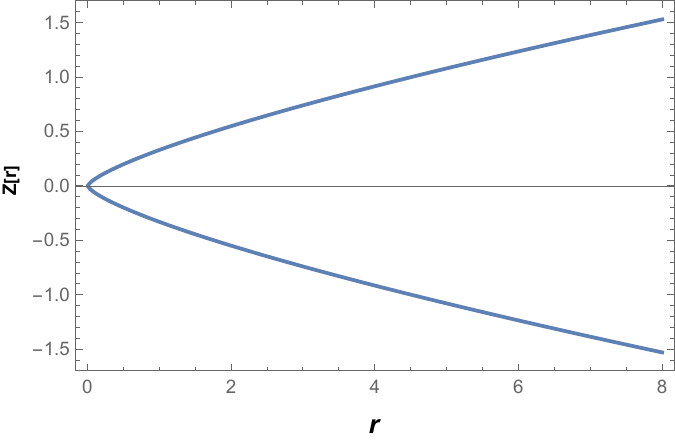}}
\subfloat[]{\includegraphics[width=0.5\linewidth]{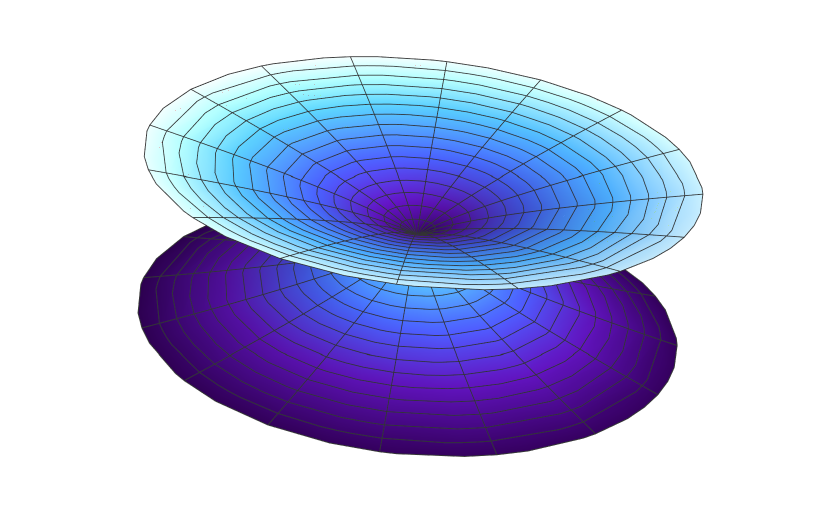}}
\caption{Embedding diagram for the shape function $b_1(r)$. The left panel shows the two-dimensional embedding profile $z(r)$, while the right panel depicts the corresponding three-dimensional embedding surface, illustrating a smooth throat and outward flaring geometry.}
\label{embedding_b1}
\end{figure*}

\begin{figure*}[htbp]
\centering
\subfloat[]{\includegraphics[width=0.45\linewidth]{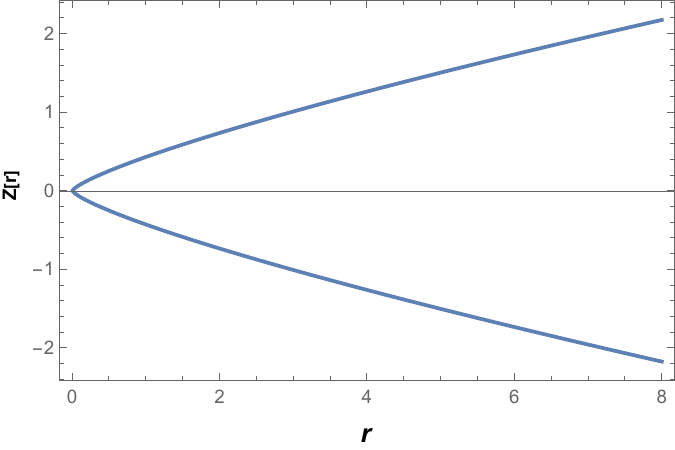}}
\subfloat[]{\includegraphics[width=0.5\linewidth]{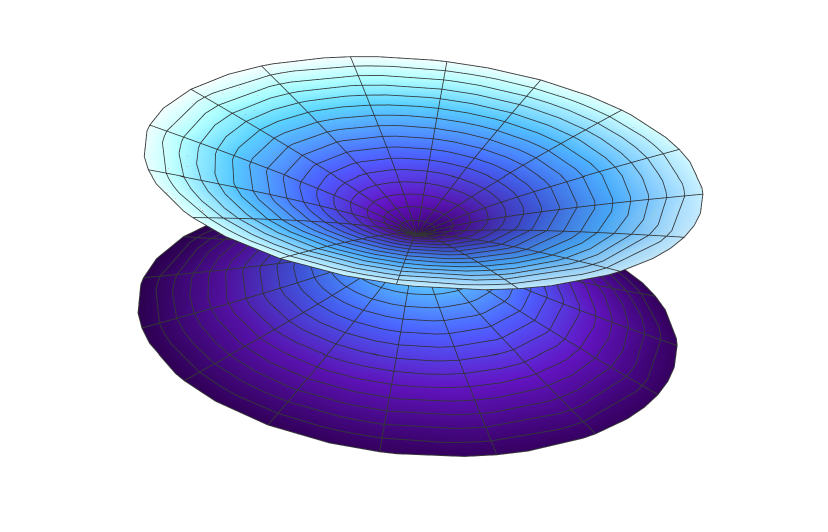}}
\caption{Embedding diagram for the shape function $b_2(r)$. The embedding surface exhibits a well-defined throat and satisfies the flaring-out condition, confirming the traversable nature of the wormhole geometry.}
\label{embedding_b2}
\end{figure*}

\begin{figure*}[htbp]
\centering
\subfloat[]{\includegraphics[width=0.45\linewidth]{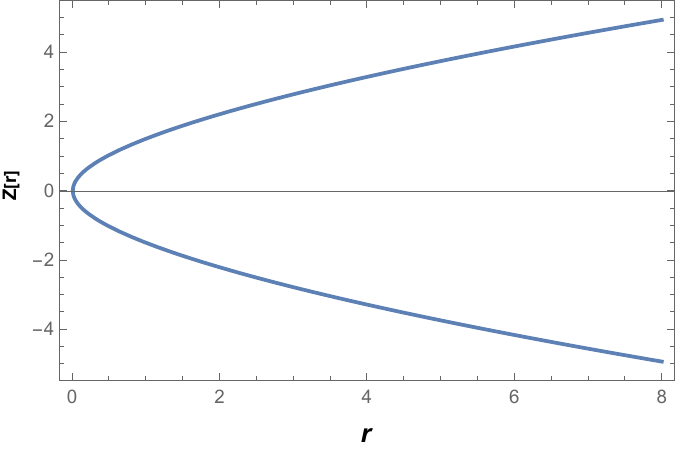}}
\subfloat[]{\includegraphics[width=0.5\linewidth]{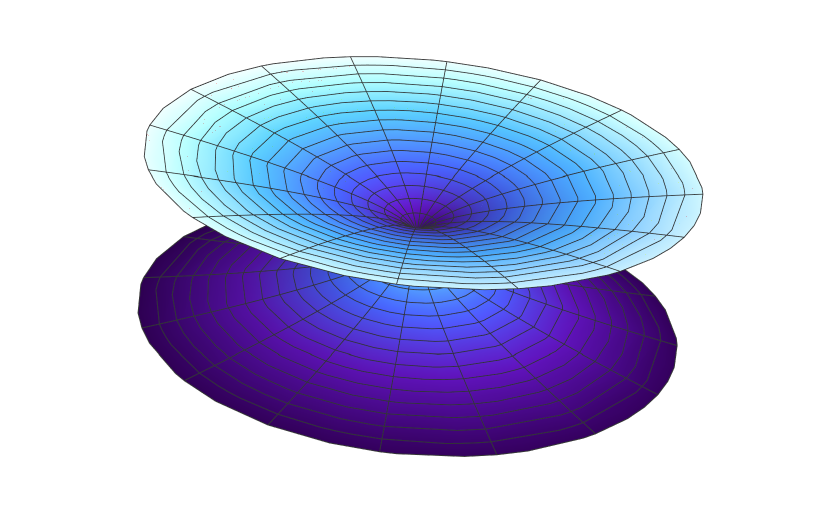}}
\caption{Embedding diagram for the shape function $b_3(r)$. The geometry smoothly opens away from the throat and approaches a nearly flat configuration at large radial distances.}
\label{embedding_b3}
\end{figure*}

From the embedding diagrams, it is evident that all three shape functions lead to physically admissible wormhole geometries. The two-dimensional profiles clearly demonstrate the outward flaring of the throat, while the three-dimensional embedding surfaces provide an intuitive geometric picture of the wormhole structure. In each case, the embedding surface is smooth at the throat and gradually flattens as $r$ increases, indicating asymptotic flatness of the spacetime.

These results confirm that the chosen shape functions satisfy the essential geometric requirements for traversable wormholes within the framework of $f(\mathscr{Q},\mathscr{L}_m)$ gravity. The embedding analysis thus complements the analytical verification of the wormhole conditions and offers a transparent geometric interpretation of the resulting spacetime configurations.

\section{Discussion}\label{section5}

Our investigation of traversable wormhole solutions in $f(\mathscr{Q},\mathscr{L}_m)$ gravity reveals several noteworthy features regarding both the geometric structure and matter content of these hypothetical spacetime tunnels. The analysis demonstrates that viable wormhole configurations can be constructed within this modified gravity framework, with the non-minimal matter-geometry coupling playing a significant role in shaping both the geometry and physical properties.

The adoption of a linear functional form $f(\mathscr{Q},\mathscr{L}_m) = -\alpha \mathscr{Q} + 2\mathscr{L}_m + \beta$ provides a tractable framework while retaining essential features of the theory. This choice represents a minimal extension of the symmetric teleparallel equivalent of general relativity, where $\alpha$ and $\beta$ serve as coupling constants that modify the effective gravitational dynamics. The linear dependence on $\mathscr{Q}$ ensures recovery of standard teleparallel gravity in appropriate limits, while the explicit coupling to $\mathscr{L}_m$ introduces novel interactions between geometry and matter.

The geometric viability of all considered shape functions—$b(r)=\sqrt{r_0 r}$, $b(r)=r_0(r/r_0)^\gamma$, and $b(r)=r_0 \ln(r+1)/\ln(r_0+1)$—is particularly significant. Each satisfies the essential traversability conditions: the flaring-out requirement $b'(r_0) < 1$, asymptotic flatness $b(r)/r \to 0$ as $r \to \infty$, and the absence of event horizons $1 - b(r)/r > 0$. This demonstrates the flexibility of $f(\mathscr{Q},\mathscr{L}_m)$ gravity in accommodating diverse geometric profiles while maintaining physical viability. The square root shape function exhibits particularly smooth geometric behavior, while the power-law and logarithmic forms offer alternative parameterizations that may be suitable for different physical scenarios.

The energy condition analysis reveals the expected violation of the null energy condition in the throat region, consistent with established wormhole physics. This violation, evident in Figure~\ref{fig:energy_conditions}, confirms that exotic matter remains necessary to sustain traversable wormholes even in $f(\mathscr{Q},\mathscr{L}_m)$ gravity. However, the degree and extent of this violation vary among the different shape functions, suggesting that careful geometric design could potentially minimize the amount of exotic matter required. The behavior of the strong and dominant energy conditions further distinguishes the various configurations, with some showing partial satisfaction of these conditions away from the throat.

The embedding diagrams presented in this work offer an intuitive geometric interpretation of the obtained wormhole solutions. The two-dimensional embedding profiles clearly exhibit the characteristic outward flaring at the throat, while the corresponding three-dimensional surfaces provide a visual representation of the wormhole structure as a smooth bridge between two asymptotically flat regions. The absence of any pinching or singular behavior in the embedding surfaces confirms that the geometric requirements for traversability are satisfied for all the chosen shape functions.

Notably, although the three shape functions differ in their analytical forms, their embedding geometries share common qualitative features, such as smooth throats and gradual flattening at large radial distances. This indicates that the non-minimal matter–geometry coupling in $f(\mathscr{Q},\mathscr{L}_m)$gravity can support a variety of wormhole geometries without introducing geometric instabilities. The embedding analysis therefore complements the energy-condition study by demonstrating that the wormhole configurations are not only mathematically consistent but also geometrically well-behaved.

The non-conservation of the energy-momentum tensor, inherent to $f(\mathscr{Q},\mathscr{L}_m)$ gravity, introduces additional complexity to the matter sector. This feature represents a fundamental departure from general relativity and could have implications for energy transport mechanisms within the wormhole. The energy exchange between geometry and matter sectors, quantified by the term $B_\nu$ in Equation~\eqref{NonCons}, may provide novel mechanisms for stabilizing wormhole geometries or mitigating energy condition violations.

The vanishing redshift function $\Phi(r)=0$ simplifies the analysis while ensuring zero tidal forces—a practical requirement for human traversability. This choice corresponds to a wormhole with no gravitational redshift, which while idealizing certain aspects, provides a clear baseline for studying geometric and matter properties. Future investigations could explore non-trivial redshift functions to understand how gravitational effects modify the solutions.

Comparative analysis with wormhole solutions in other modified gravity theories suggests that $f(\mathscr{Q},\mathscr{L}_m)$ gravity offers a distinct geometric foundation while sharing some qualitative features with curvature-based alternatives. The teleparallel framework, built on non-metricity rather than curvature, provides a mathematically distinct but physically rich setting for exploring exotic geometries.

\section{Conclusion}\label{section6}

This work has systematically investigated traversable wormhole solutions within the framework of $f(\mathscr{Q},\mathscr{L}_m)$ gravity, a symmetric teleparallel theory featuring non-minimal coupling between geometry and matter. By adopting a linear functional form $f(\mathscr{Q},\mathscr{L}_m) = -\alpha \mathscr{Q} + 2\mathscr{L}_m + \beta$, we derived the field equations for static, spherically symmetric Morris-Thorne wormholes with vanishing redshift function. Three distinct shape functions were analyzed for their geometric viability and energy condition profiles.

The principal findings of this study are:
\begin{enumerate}
    \item \textbf{Geometric Viability}: All considered shape functions satisfy the essential traversability conditions—flaring-out requirement, asymptotic flatness, and absence of event horizons—confirming that $f(\mathscr{Q},\mathscr{L}_m)$ gravity supports diverse wormhole geometries. Furthermore, the embedding diagram analysis provides a clear geometric visualization of the wormhole spacetimes supported by $f(\mathscr{Q},\mathscr{L}_m)$ gravity. The smooth, outward-flaring embedding surfaces confirm the presence of well-defined throats and asymptotically flat regions for all considered shape functions. This geometric perspective strengthens the physical interpretation of the solutions and reinforces the viability of symmetric teleparallel gravity with matter–geometry coupling as a framework for traversable wormholes.

    \item \textbf{Energy Condition Violation}: Consistent with fundamental wormhole physics, the null energy condition is violated near the throat in all cases, indicating the necessity of exotic matter. The extent and profile of this violation depend on the specific shape function.

    \item \textbf{Matter-Geometry Coupling}: The non-minimal coupling in $f(\mathscr{Q},\mathscr{L}_m)$ gravity introduces novel interactions between the geometric and matter sectors, manifesting in non-conservation of the energy-momentum tensor and modified gravitational dynamics.

    \item \textbf{Parameter Dependence}: The coupling constants $\alpha$ and $\beta$ influence both geometric and physical properties, offering tunable parameters for constructing wormholes with desired characteristics.
\end{enumerate}

The results demonstrate that $f(\mathscr{Q},\mathscr{L}_m)$ gravity provides a viable theoretical framework for constructing traversable wormholes, with the non-minimal matter-geometry coupling offering additional degrees of freedom compared to standard general relativity. While exotic matter remains necessary, the theory's flexible structure may enable configurations that minimize or localize energy condition violations.

Several promising directions for future work include investigating nonlinear $f(\mathscr{Q},\mathscr{L}_m)$ forms and nontrivial redshift functions to better model gravitational effects, performing stability analyses under perturbations, exploring observational signatures like gravitational lensing to distinguish these wormholes from other compact objects, and examining dynamic wormholes in cosmological contexts.

\section*{Acknowledgments}

CSV and VV acknowledge DST, New Delhi, India, for supporting research facilities under DST-FIST-2019.

\section*{Data Availability}

No new data were generated or analyzed in this study.

\bibliography{Ref.bib}

@article{morris1988wormholes,
  title="{Wormholes in spacetime and their use for interstellar travel: A tool for teaching general relativity}",
  author="Morris, Michael S and Thorne, Kip S",
  journal="American Journal of Physics",
  volume="56",
  number="5",
  pages="395-412",
  year="1988",
  publisher="American Association of Physics Teachers"
}

@article{Kavya:2025glb,
    author = "Kavya, N. S. and Varsha, C. S. and Sudharani, L. and Venkatesha, V.",
    title = "{Unifying non-commutative geometry with Casimir energy: A novel f(R) wormhole solution}",
    doi = "10.1016/j.nuclphysb.2025.116794",
    journal = "Nucl. Phys. B",
    volume = "1011",
    pages = "116794",
    year = "2025"
}

@book{Visser:1995cc,
    author = "Visser, Matt",
    title = "{Lorentzian wormholes: From Einstein to Hawking}",
    isbn = "978-1-56396-653-8",
    year = "1995",
    publisher ="American Inst. of Physics, "
}

@article{Venkatesha:2023zkm,
    author = "Venkatesha, V. and Kavya, N. S. and Sahoo, P. K.",
    title = "{Geometric structures of Morris-Thorne wormhole metric in f(, \ensuremath{\mathscr{L}}$_{m}$) gravity and energy conditions}",
    eprint = "2305.04707",
    archivePrefix = "arXiv",
    primaryClass = "gr-qc",
    doi = "10.1088/1402-4896/acd483",
    journal = "Phys. Scripta",
    volume = "98",
    number = "6",
    pages = "065020",
    year = "2023"
}

@article{Gao:2016bin,
    author = "Gao, Ping and Jafferis, Daniel Louis and Wall, Aron C.",
    title = "{Traversable Wormholes via a Double Trace Deformation}",
    eprint = "1608.05687",
    archivePrefix = "arXiv",
    primaryClass = "hep-th",
    doi = "10.1007/JHEP12(2017)151",
    journal = "JHEP",
    volume = "12",
    pages = "151",
    year = "2017"
}

@article{Kavya:2023tjf,
    author = "Kavya, N. S. and Venkatesha, V. and Mustafa, G. and Sahoo, P. K. and Rashmi, S. V. Divya",
    title = "{Static traversable wormhole solutions in f(R,\ensuremath{\mathscr{L}}m) gravity}",
    eprint = "2305.01469",
    archivePrefix = "arXiv",
    primaryClass = "gr-qc",
    doi = "10.1016/j.cjph.2023.05.002",
    journal = "Chin. J. Phys.",
    volume = "84",
    pages = "1-11",
    year = "2023"
}

@article{Armendariz-Picon:2002gjc,
    author = "Armendariz-Picon, C.",
    title = "{On a class of stable, traversable Lorentzian wormholes in classical general relativity}",
    eprint = "gr-qc/0201027",
    archivePrefix = "arXiv",
    doi = "10.1103/PhysRevD.65.104010",
    journal = "Phys. Rev. D",
    volume = "65",
    pages = "104010",
    year = "2002"
}

@article{Nicolis:2009qm,
    author = "Nicolis, Alberto and Rattazzi, Riccardo and Trincherini, Enrico",
    title = "{Energy's and amplitudes' positivity}",
    eprint = "0912.4258",
    archivePrefix = "arXiv",
    primaryClass = "hep-th",
    doi = "10.1007/JHEP05(2010)095",
    journal = "JHEP",
    volume = "05",
    pages = "095",
    year = "2010",
    note = "[Erratum: JHEP 11, 128 (2011)]"
}

@article{Kuhfittig:1999nd,
    author = "Kuhfittig, P. K. F.",
    title = "{A wormhole with a special shape function}",
    doi = "10.1119/1.19206",
    journal = "Am. J. Phys.",
    volume = "67",
    pages = "125--126",
    year = "1999"
}

@article{Lobo:2009ip,
    author = "Lobo, Francisco S. N. and Oliveira, Miguel A.",
    title = "{Wormhole geometries in f(R) modified theories of gravity}",
    eprint = "0909.5539",
    archivePrefix = "arXiv",
    primaryClass = "gr-qc",
    doi = "10.1103/PhysRevD.80.104012",
    journal = "Phys. Rev. D",
    volume = "80",
    pages = "104012",
    year = "2009"
}

@article{Lobo:2020kxn,
    author = "Lobo, Francisco S. N. and Simpson, Alex and Visser, Matt",
    title = "{Dynamic thin-shell black-bounce traversable wormholes}",
    eprint = "2003.09419",
    archivePrefix = "arXiv",
    primaryClass = "gr-qc",
    doi = "10.1103/PhysRevD.101.124035",
    journal = "Phys. Rev. D",
    volume = "101",
    number = "12",
    pages = "124035",
    year = "2020"
}

@article{Harko:2013yb,
    author = "Harko, Tiberiu and Lobo, Francisco S. N. and Mak, M. K. and Sushkov, Sergey V.",
    title = "{Modified-gravity wormholes without exotic matter}",
    eprint = "1301.6878",
    archivePrefix = "arXiv",
    primaryClass = "gr-qc",
    doi = "10.1103/PhysRevD.87.067504",
    journal = "Phys. Rev. D",
    volume = "87",
    number = "6",
    pages = "067504",
    year = "2013"
}

@article{Kanti:2011jz,
    author = "Kanti, Panagiota and Kleihaus, Burkhard and Kunz, Jutta",
    title = "{Wormholes in Dilatonic Einstein-Gauss-Bonnet Theory}",
    eprint = "1108.3003",
    archivePrefix = "arXiv",
    primaryClass = "gr-qc",
    doi = "10.1103/PhysRevLett.107.271101",
    journal = "Phys. Rev. Lett.",
    volume = "107",
    pages = "271101",
    year = "2011"
}

@article{Usmani:2010cd,
    author = "Usmani, A. A. and Hasan, Z. and Rahaman, F. and Rakib, Sk. A. and Ray, Saibal and Kuhfittig, Peter K. F.",
    title = "{Thin-shell wormholes from charged black holes in generalized dilaton-axion gravity}",
    eprint = "1001.1415",
    archivePrefix = "arXiv",
    primaryClass = "gr-qc",
    doi = "10.1007/s10714-010-1044-y",
    journal = "Gen. Rel. Grav.",
    volume = "42",
    pages = "2901--2912",
    year = "2010"
}

@article{Zubair:2016cde,
    author = "Zubair, M. and Waheed, Saira and Ahmad, Yasir",
    title = "{Static spherically symmetric wormholes in f(R, T) gravity}",
    eprint = "1607.05998",
    archivePrefix = "arXiv",
    primaryClass = "gr-qc",
    doi = "10.1140/epjc/s10052-016-4288-1",
    journal = "Eur. Phys. J. C",
    volume = "76",
    number = "8",
    pages = "444",
    year = "2016"
}

@article{Ovgun:2018fnk,
    author = {\"Ovg\"un, Ali},
    title = "{Light deflection by Damour-Solodukhin wormholes and Gauss-Bonnet theorem}",
    eprint = "1805.06296",
    archivePrefix = "arXiv",
    primaryClass = "gr-qc",
    doi = "10.1103/PhysRevD.98.044033",
    journal = "Phys. Rev. D",
    volume = "98",
    number = "4",
    pages = "044033",
    year = "2018"
}

@article{Kuhfittig:2013hva,
    author = "Kuhfittig, Peter K. F.",
    title = "{Gravitational lensing of wormholes in the galactic halo region}",
    eprint = "1311.2274",
    archivePrefix = "arXiv",
    primaryClass = "gr-qc",
    doi = "10.1140/epjc/s10052-014-2818-2",
    journal = "Eur. Phys. J. C",
    volume = "74",
    number = "99",
    pages = "2818",
    year = "2014"
}

@article{Kuhfittig:2005gy,
    author = "Kuhfittig, Peter K. F.",
    title = "{More on wormholes supported by small amounts of exotic matter}",
    eprint = "gr-qc/0512027",
    archivePrefix = "arXiv",
    doi = "10.1103/PhysRevD.73.084014",
    journal = "Phys. Rev. D",
    volume = "73",
    pages = "084014",
    year = "2006"
}

@article{Hassan:2022hcb,
    author = "Hassan, Zinnat and Ghosh, Sayantan and Sahoo, P. K. and Bamba, Kazuharu",
    title = "{Casimir wormholes in modified symmetric teleparallel gravity}",
    eprint = "2207.09945",
    archivePrefix = "arXiv",
    primaryClass = "gr-qc",
    doi = "10.1140/epjc/s10052-022-11107-0",
    journal = "Eur. Phys. J. C",
    volume = "82",
    number = "12",
    pages = "1116",
    year = "2022"
}

@article{Mishra:2024shg,
    author = "Mishra, Sai Swagat and Kavya, N. S. and Sahoo, P. K. and Venkatesha, V.",
    title = "{Chebyshev cosmography in the framework of extended symmetric teleparallel theory}",
    eprint = "2412.03065",
    archivePrefix = "arXiv",
    primaryClass = "gr-qc",
    doi = "10.1016/j.dark.2024.101759",
    journal = "Phys. Dark Univ.",
    volume = "47",
    pages = "101759",
    year = "2025"
}

@article{Mishra:2024oln,
    author = "Mishra, Sai Swagat and Kavya, N. S. and Sahoo, P. K. and Venkatesha, V.",
    title = "{Constraining Extended Teleparallel Gravity via Cosmography: A Model-independent Approach}",
    eprint = "2406.06661",
    archivePrefix = "arXiv",
    primaryClass = "gr-qc",
    doi = "10.3847/1538-4357/ad5555",
    journal = "Astrophys. J.",
    volume = "970",
    number = "1",
    pages = "57",
    year = "2024"
}

@article{Kavya:2023mwv,
    author = "Kavya, N. S. and Venkatesha, V. and Mustafa, G. and Sahoo, P. K.",
    title = "{On possible wormhole solutions supported by non-commutative geometry within f(R,\ensuremath{\mathscr{L}}m) gravity}",
    eprint = "2307.02498",
    archivePrefix = "arXiv",
    primaryClass = "gr-qc",
    doi = "10.1016/j.aop.2023.169383",
    journal = "Annals Phys.",
    volume = "455",
    pages = "169383",
    year = "2023"
}

@article{Capozziello:2022wgl,
    author = "Capozziello, Salvatore and D'Agostino, Rocco",
    title = "{Model-independent reconstruction of f(Q) non-metric gravity}",
    eprint = "2204.01015",
    archivePrefix = "arXiv",
    primaryClass = "gr-qc",
    doi = "10.1016/j.physletb.2022.137229",
    journal = "Phys. Lett. B",
    volume = "832",
    pages = "137229",
    year = "2022"
}

@article{Lazkoz:2019sjl,
    author = "Lazkoz, Ruth and Lobo, Francisco S. N. and Ortiz-Ba\~nos, Mar\'\i{}a and Salzano, Vincenzo",
    title = "{Observational constraints of $f(Q)$ gravity}",
    eprint = "1907.13219",
    archivePrefix = "arXiv",
    primaryClass = "gr-qc",
    doi = "10.1103/PhysRevD.100.104027",
    journal = "Phys. Rev. D",
    volume = "100",
    number = "10",
    pages = "104027",
    year = "2019"
}

@article{Anagnostopoulos:2021ydo,
    author = "Anagnostopoulos, Fotios K. and Basilakos, Spyros and Saridakis, Emmanuel N.",
    title = "{First evidence that non-metricity f(Q) gravity could challenge \ensuremath{\Lambda}CDM}",
    eprint = "2104.15123",
    archivePrefix = "arXiv",
    primaryClass = "gr-qc",
    doi = "10.1016/j.physletb.2021.136634",
    journal = "Phys. Lett. B",
    volume = "822",
    pages = "136634",
    year = "2021"
}

@article{Mustafa:2021ykn,
    author = "Mustafa, G. and Hassan, Zinnat and Moraes, P. H. R. S. and Sahoo, P. K.",
    title = "{Wormhole solutions in symmetric teleparallel gravity}",
    eprint = "2108.01446",
    archivePrefix = "arXiv",
    primaryClass = "gr-qc",
    doi = "10.1016/j.physletb.2021.136612",
    journal = "Phys. Lett. B",
    volume = "821",
    pages = "136612",
    year = "2021"
}

@article{Kavya:2024ssu,
    author = "Kavya, N. S. and Mishra, Sai Swagat and Sahoo, P. K. and Venkatesha, V.",
    title = "{Can teleparallel f(T) models play a bridge between early and late time Universe?}",
    eprint = "2407.09589",
    archivePrefix = "arXiv",
    primaryClass = "gr-qc",
    doi = "10.1093/mnras/stae1723",
    journal = "Mon. Not. Roy. Astron. Soc.",
    volume = "532",
    number = "3",
    pages = "3126--3133",
    year = "2024"
}

@article{Kavya:2024bpj,
    author = "Kavya, N. S. and Venkatesha, V.",
    title = "{Embedding the \ensuremath{\Lambda}CDM framework in non-minimal f(Q) gravity with matter-coupling}",
    doi = "10.1016/j.physletb.2024.138927",
    journal = "Phys. Lett. B",
    volume = "856",
    pages = "138927",
    year = "2024"
}

@article{Xu:2019sbp,
    author = "Xu, Yixin and Li, Guangjie and Harko, Tiberiu and Liang, Shi-Dong",
    title = "{$f(Q,T)$ gravity}",
    eprint = "1908.04760",
    archivePrefix = "arXiv",
    primaryClass = "gr-qc",
    doi = "10.1140/epjc/s10052-019-7207-4",
    journal = "Eur. Phys. J. C",
    volume = "79",
    number = "8",
    pages = "708",
    year = "2019"
}

@article{BeltranJimenez:2017tkd,
    author = "Beltr\'an Jim\'enez, Jose and Heisenberg, Lavinia and Koivisto, Tomi",
    title = "{Coincident General Relativity}",
    eprint = "1710.03116",
    archivePrefix = "arXiv",
    primaryClass = "gr-qc",
    reportNumber = "NORDITA-2017-100, IFT-UAM/CSIC-17-093, ITS-ETH-2017-10",
    doi = "10.1103/PhysRevD.98.044048",
    journal = "Phys. Rev. D",
    volume = "98",
    number = "4",
    pages = "044048",
    year = "2018"
}

@article{Naik:2023ykt,
    author = "Naik, Devaraja Mallesha and Kavya, N. S. and Sudharani, L. and Venkatesha, V.",
    title = "{Impact of a newly parametrized deceleration parameter on the accelerating universe and the reconstruction of f(Q) non-metric gravity models}",
    doi = "10.1140/epjc/s10052-023-12029-1",
    journal = "Eur. Phys. J. C",
    volume = "83",
    number = "9",
    pages = "840",
    year = "2023"
}

@article{Najera:2021puo,
    author = "N\'ajera, Antonio and Fajardo, Amanda",
    title = "{Fitting f(Q,T) gravity models with a \ensuremath{\Lambda}CDM limit using H(z) and Pantheon data}",
    eprint = "2104.14065",
    archivePrefix = "arXiv",
    primaryClass = "gr-qc",
    doi = "10.1016/j.dark.2021.100889",
    journal = "Phys. Dark Univ.",
    volume = "34",
    pages = "100889",
    year = "2021"
}

@article{Shiravand:2022ccb,
    author = "Shiravand, Maryam and Fakhry, Saeed and Farhoudi, Mehrdad",
    title = "{Cosmological inflation in f(Q,T) gravity}",
    eprint = "2204.00906",
    archivePrefix = "arXiv",
    primaryClass = "gr-qc",
    doi = "10.1016/j.dark.2022.101106",
    journal = "Phys. Dark Univ.",
    volume = "37",
    pages = "101106",
    year = "2022"
}

@article{Tayde:2022lxd,
    author = "Tayde, Moreshwar and Hassan, Zinnat and Sahoo, P. K. and Gutti, Sashideep",
    title = "{Static spherically symmetric wormholes in gravity*}",
    eprint = "2206.01184",
    archivePrefix = "arXiv",
    primaryClass = "gr-qc",
    doi = "10.1088/1674-1137/ac7f22",
    journal = "Chin. Phys. C",
    volume = "46",
    number = "11",
    pages = "115101",
    year = "2022"
}

@article{Boehmer:2007md,
    author = "Boehmer, Christian G. and Harko, Tiberiu and Lobo, Francisco S. N.",
    title = "{Wormhole geometries with conformal motions}",
    eprint = "0711.2424",
    archivePrefix = "arXiv",
    primaryClass = "gr-qc",
    doi = "10.1088/0264-9381/25/7/075016",
    journal = "Class. Quant. Grav.",
    volume = "25",
    pages = "075016",
    year = "2008"
}

@article{Boehmer:2007rm,
    author = "Boehmer, Christian G. and Harko, Tiberiu and Lobo, Francisco S. N.",
    title = "{Conformally symmetric traversable wormholes}",
    eprint = "0708.1537",
    archivePrefix = "arXiv",
    primaryClass = "gr-qc",
    doi = "10.1103/PhysRevD.76.084014",
    journal = "Phys. Rev. D",
    volume = "76",
    pages = "084014",
    year = "2007"
}

@article{Kavya:2023lms,
    author = "Kavya, N. S. and Mustafa, G. and Venkatesha, V. and Sahoo, P. K.",
    title = "{Exploring wormhole solutions in curvature\textendash{}matter coupling gravity supported by noncommutative geometry and conformal symmetry}",
    eprint = "2306.08856",
    archivePrefix = "arXiv",
    primaryClass = "gr-qc",
    doi = "10.1016/j.cjph.2024.01.004",
    journal = "Chin. J. Phys.",
    volume = "87",
    pages = "751--765",
    year = "2024"
}

@article{Sushkov:2005kj,
    author = "Sushkov, Sergey V.",
    title = "{Wormholes supported by a phantom energy}",
    eprint = "gr-qc/0502084",
    archivePrefix = "arXiv",
    doi = "10.1103/PhysRevD.71.043520",
    journal = "Phys. Rev. D",
    volume = "71",
    pages = "043520",
    year = "2005"
}

@article{Mustafa:2019oiy,
    author = "Mustafa, G. and Shahzad, M. R. and Abbas, G. and Xia, T.",
    title = "{Stable wormholes solutions in the background of Rastall theory}",
    doi = "10.1142/S0217732320500352",
    journal = "Mod. Phys. Lett. A",
    volume = "35",
    number = "07",
    pages = "2050035",
    year = "2019"
}

@article{Banerjee:2019wjj,
    author = "Banerjee, Ayan and Singh, Ksh. Newton and Jasim, M. K. and Rahaman, Farook",
    title = "{Conformally symmetric traversable wormholes in $f(R,T)$ gravity}",
    eprint = "1908.04754",
    archivePrefix = "arXiv",
    primaryClass = "gr-qc",
    doi = "10.1016/j.aop.2020.168295",
    journal = "Annals Phys.",
    volume = "422",
    pages = "168295",
    year = "2020"
}

@article{SinghKsh:2023fphy,
    author = "Ksh. Newton SinghKsh. Farook Rahaman, Debabrata Deb and S. K. Maurya.",
    title = "{Traversable Finslerian wormholes supported by phantom energy}",
    eprint = "  ",
    archivePrefix = " ",
    primaryClass = " ",
    doi = "10.3389/fphy.2022.1038905",
    journal = "Frontiers in Physics.",
    volume = " 10",
    pages = "1038905",
    year = "2023"
}

@article{Bhar:2023ihi,
    author = "Bhar, Piyali and Rej, Pramit and Singh, Ksh. Newton",
    title = "{New classes of wormhole model in f(R,T) gravity by assuming conformal motion}",
    doi = "10.1016/j.newast.2023.102059",
    journal = "New Astron.",
    volume = "103",
    pages = "102059",
    year = "2023"
}

@article{Banerjee:2021mqk,
    author = "Banerjee, Ayan and Pradhan, Anirudh and Tangphati, Takol and Rahaman, Farook",
    title = "{Wormhole geometries in $f(Q)$ gravity and the energy conditions}",
    eprint = "2109.15105",
    archivePrefix = "arXiv",
    primaryClass = "gr-qc",
    doi = "10.1140/epjc/s10052-021-09854-7",
    journal = "Eur. Phys. J. C",
    volume = "81",
    number = "11",
    pages = "1031",
    year = "2021"
}

@article{Varsha:2025qgv,
    author = "Varsha, C. S. and Sudharani, L. and Kavya, N. S. and Venkatesha, V.",
    title = "{Thermodynamic insights into evolving Lorentzian wormholes in f(R,T) gravity}",
    doi = "10.1016/j.physletb.2025.139590",
    journal = "Phys. Lett. B",
    volume = "867",
    pages = "139590",
    year = "2025"
}

@article{Varsha:2025oik,
    author = "Varsha, C. S. and Sudharani, L. and Kavya, N. S. and Venkatesha, V.",
    title = "{A novel theoretical approach to dark matter supported Casimir wormhole}",
    doi = "10.1016/j.nuclphysb.2025.116929",
    journal = "Nucl. Phys. B",
    volume = "1017",
    pages = "116929",
    year = "2025"
}

@article{Varsha:2024iwx,
    author = "Varsha, C. S. and Sudharani, L. and Kavya, N. S. and Venkatesha, V.",
    title = "{Testing Curvature-Matter Coupling Gravity via Swampland Conjectures}",
    doi = "10.1002/prop.202400160",
    journal = "Fortsch. Phys.",
    volume = "73",
    number = "5",
    pages = "2400160",
    year = "2025"
}

@article{venkatesha2025noncommutative,
  title={Noncommutative wormholes and conformal symmetry: a study in $ f (Q, T) $ gravity},
  author={Venkatesha, V and Lathakumari, GN and Varsha, CS},
  journal={Indian Journal of Physics},
  pages={1--13},
  year={2025},
  publisher={Springer}
}

@article{Kavya:2022dam,
    author = "Kavya, N. S. and Venkatesha, V. and Mandal, Sanjay and Sahoo, P. K.",
    title = "{Constraining anisotropic cosmological model in f(R,{\ensuremath{\mathscr{L}}}m) Gravity}",
    eprint = "2210.09307",
    archivePrefix = "arXiv",
    primaryClass = "gr-qc",
    doi = "10.1016/j.dark.2022.101126",
    journal = "Phys. Dark Univ.",
    volume = "38",
    pages = "101126",
    year = "2022"
}

@article{Venkatesha:2023tay,
    author = "Venkatesha, V. and Chalavadi, Chaitra Chooda and Kavya, N. S. and Sahoo, P. K.",
    title = "{Wormhole geometry and three-dimensional embedding in extended symmetric teleparallel gravity}",
    eprint = "2308.07862",
    archivePrefix = "arXiv",
    primaryClass = "gr-qc",
    doi = "10.1016/j.newast.2023.102090",
    journal = "New Astron.",
    volume = "105",
    pages = "102090",
    year = "2024"
}

@article{Chalavadi:2023zcw,
    author = "Chalavadi, Chaitra Chooda and Kavya, N. S. and Venkatesha, V.",
    title = "{Wormhole solutions supported by non-commutative geometric background in $f (\mathcal {Q},\mathcal {T})$ gravity}",
    doi = "10.1140/epjp/s13360-023-04480-6",
    journal = "Eur. Phys. J. Plus",
    volume = "138",
    number = "10",
    pages = "885",
    year = "2023"
}

@mastersthesis{Kavya:2025ygq,
    author = "Kavya, N. S.",
    title = "{The Study on Modified Theories of General Relativity: A Differential Geometric Approach}",
    eprint = "2507.04031",
    archivePrefix = "arXiv",
    primaryClass = "gr-qc",
    type = "Other thesis",
    month = "7",
    year = "2025"
}

@article{Rahaman:2013xoa,
    author = "Rahaman, Farook and Kuhfittig, P. K. F. and Ray, Saibal and Islam, Nasarul",
    title = "{Possible existence of wormholes in the galactic halo region}",
    eprint = "1307.1237",
    archivePrefix = "arXiv",
    primaryClass = "gr-qc",
    doi = "10.1140/epjc/s10052-014-2750-5",
    journal = "Eur. Phys. J. C",
    volume = "74",
    pages = "2750",
    year = "2014"
}

@article{Rahaman:2006vg,
    author = "Rahaman, F. and Kalam, M. and Chakraborty, S.",
    title = "{Thin shell wormholes in higher dimensiaonal Einstein-Maxwell theory}",
    eprint = "gr-qc/0607061",
    archivePrefix = "arXiv",
    doi = "10.1007/s10714-006-0325-y",
    journal = "Gen. Rel. Grav.",
    volume = "38",
    pages = "1687--1695",
    year = "2006"
}

@article{Rahaman:2010bt,
    author = "Rahaman, Farook and Kuhfittig, P. K. F. and Kalam, M. and Usmani, A. A. and Ray, Saibal",
    title = "{A comparison of Ho{\v{r}}ava-Lifshitz gravity and Einstein gravity through thin-shell wormhole construction}",
    eprint = "1011.3600",
    archivePrefix = "arXiv",
    primaryClass = "gr-qc",
    doi = "10.1088/0264-9381/28/15/155021",
    journal = "Class. Quant. Grav.",
    volume = "28",
    pages = "155021",
    year = "2011"
}

@article{Jamil:2013tva,
    author = "Jamil, Mubasher and Rahaman, Farook and Myrzakulov, Ratbay and Kuhfittig, P. K. F. and Ahmed, Nasr and Mondal, Umar F",
    title = "{Nonommutative wormholes in $f(R)$ gravity}",
    eprint = "1304.2240",
    archivePrefix = "arXiv",
    primaryClass = "gr-qc",
    doi = "10.3938/jkps.65.917",
    journal = "J. Korean Phys. Soc.",
    volume = "65",
    number = "6",
    pages = "917--925",
    year = "2014"
}

@article{Rahaman:2013qza,
    author = "Rahaman, Farook and Banerjee, Ayan and Jamil, Mubasher and Yadav, Anil Kumar and Idris, Humaira",
    title = "{Noncommutative Wormholes in f(R) Gravity with Lorentzian Distribution}",
    eprint = "1312.7684",
    archivePrefix = "arXiv",
    primaryClass = "gr-qc",
    doi = "10.1007/s10773-013-1993-5",
    journal = "Int. J. Theor. Phys.",
    volume = "53",
    pages = "1910--1919",
    year = "2014"
}

@article{Hazarika:2024alm,
    author = "Hazarika, Ayush and Arora, Simran and Sahoo, P. K. and Harko, Tiberiu",
    title = "{f(Q,Lm) gravity, and its cosmological implications}",
    eprint = "2407.00989",
    archivePrefix = "arXiv",
    primaryClass = "gr-qc",
    doi = "10.1016/j.dark.2025.102092",
    journal = "Phys. Dark Univ.",
    volume = "50",
    pages = "102092",
    year = "2025"
}

@article{Hassan:2021egb,
    author = "Hassan, Zinnat and Mandal, Sanjay and Sahoo, P. K.",
    title = "{Traversable Wormhole Geometries in Gravity}",
    eprint = "2102.00915",
    archivePrefix = "arXiv",
    primaryClass = "gr-qc",
    doi = "10.1002/prop.202100023",
    journal = "Fortsch. Phys.",
    volume = "69",
    number = "6",
    pages = "2100023",
    year = "2021"
}

@article{Moraes:2019pao,
    author = "Moraes, P. H. R. S. and Sahoo, P. K.",
    title = "{Wormholes in exponential $f(R,T)$ gravity}",
    eprint = "1903.03421",
    archivePrefix = "arXiv",
    primaryClass = "gr-qc",
    doi = "10.1140/epjc/s10052-019-7206-5",
    journal = "Eur. Phys. J. C",
    volume = "79",
    number = "8",
    pages = "677",
    year = "2019"
}

@article{Venkatesha:2025cbl,
    author = "Venkatesha, V. and Lathakumari, G. N. and Varsha, C. S.",
    title = "{Noncommutative wormholes and conformal symmetry: a study in $f(Q,T)$ gravity}",
    doi = "10.1007/s12648-025-03727-5",
    journal = "Indian J. Phys.",
    volume = "99",
    number = "13",
    pages = "5301--5313",
    year = "2025"
}

@article{Myrzakulov:2024kuu,
    author = {Myrzakulov, Kairat and Koussour, M. and Donmez, O. and Cilli, A. and G{\"u}dekli, E. and Rayimbaev, J.},
    title = "{Observational analysis of late-time acceleration in f(Q,Lm) gravity}",
    eprint = "2409.18920",
    archivePrefix = "arXiv",
    primaryClass = "astro-ph.CO",
    doi = "10.1016/j.jheap.2024.09.014",
    journal = "JHEAp",
    volume = "44",
    pages = "164--171",
    year = "2024"
}

@article{Samaddar:2024qno,
    author = "Samaddar, Amit and Singh, S. Surendra",
    title = "{A novel approach to baryogenesis in f(Q,Lm) gravity and its cosmological implications}",
    eprint = "2410.05335",
    archivePrefix = "arXiv",
    primaryClass = "gr-qc",
    doi = "10.1016/j.nuclphysb.2025.116834",
    journal = "Nucl. Phys. B",
    volume = "1012",
    pages = "116834",
    year = "2025"
}

@article{Samaddar:2025qxr,
    author = "Samaddar, Amit and Surendra Singh, S.",
    title = "{Observational viability of generalized Chaplygin gas in f(Q, L{\_}{m}) gravity}",
    doi = "10.1007/s10509-025-04483-y",
    journal = "Astrophys. Space Sci.",
    volume = "370",
    number = "9",
    pages = "92",
    year = "2025"
}

@article{Myrzakulov:2025yxy,
    author = "Myrzakulov, Y. and Donmez, O. and Koussour, M. and Muminov, S. and Ostemir, D. and Rayimbaev, J.",
    title = "{Energy conditions in $f(Q, L_m)$ gravity}",
    eprint = "2504.03855",
    archivePrefix = "arXiv",
    primaryClass = "astro-ph.CO",
    doi = "10.1140/epjc/s10052-025-14112-1",
    journal = "Eur. Phys. J. C",
    volume = "85",
    number = "4",
    pages = "376",
    year = "2025"
}

@article{Myrzakulov:2025ycj,
    author = "Myrzakulov, Y. and Alfedeel, Alnadhief H. A. and Koussour, M. and Muminov, S. and Hassan, E. I. and Rayimbaev, J.",
    title = "{Modified cosmology in f(Q,Lm) gravity}",
    eprint = "2505.00208",
    archivePrefix = "arXiv",
    primaryClass = "astro-ph.CO",
    doi = "10.1016/j.physletb.2025.139506",
    journal = "Phys. Lett. B",
    volume = "866",
    pages = "139506",
    year = "2025"
}

@article{Sharif:2025rko,
    author = "Sharif, M. and Zeeshan Gul, M. and Mahmood, Rida",
    title = "{Realization of a viscous bounce in f(Q,Lm) gravity}",
    doi = "10.1016/j.hedp.2025.101235",
    journal = "High Energy Dens. Phys.",
    volume = "57",
    pages = "101235",
    year = "2025"
}

\end{document}